\documentclass[journal]{IEEEtran}
\usepackage{graphicx}
\usepackage{caption}
\usepackage{subcaption}
\usepackage{float}

\newcommand{\fig}[1]{Fig.~\ref{#1}}

\begin{document}
\title{Construction and Test of New Precision Drift-Tube Chambers for Upgrades of the ATLAS Muon Spectrometer in 2016/17}

\author{\underline{H.~Kroha$^*$}\thanks{$^*$Corresponding author: kroha@mpp.mpg.de}, O.~Kortner, F.~M\"uller, S.~Nowak, K.~Schmidt-Sommerfeld
 \\ \textit{Max-Planck-Institut f\"ur Physik, Munich}}

\maketitle
\pagestyle{empty}
\thispagestyle{empty}

\begin{abstract}
Small-diameter Muon Drift Tube (sMDT) chambers have been developed for the ATLAS muon detector upgrade. They possess an improved rate capability and a more compact design with respect to the existing chambers, which allows to equip detector regions uninstrument at present. The chamber assembly methods have been optimized for mass production, while the sense wire positioning accuracy is improved to below ten microns. The chambers will be ready for installation in the winter shutdown 2016/17 of the Large Hadron Collider. The design and construction of the new sMDT chambers for ATLAS will be discussed as well as measurements of their precision and performance.
\end{abstract}

\begin{figure}[H]
	\centering
	\begin{subfigure}[b]{0.45\textwidth}
		\centering
		\includegraphics[width=1\textwidth]{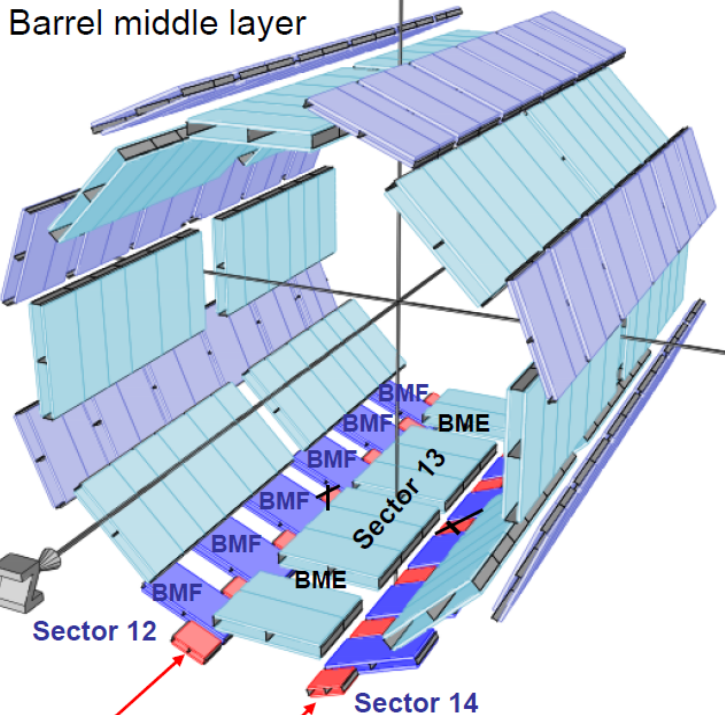}
	\end{subfigure}
	\qquad
	\begin{subfigure}[b]{0.45\textwidth}
		\centering
		\includegraphics[width=1.0\textwidth]{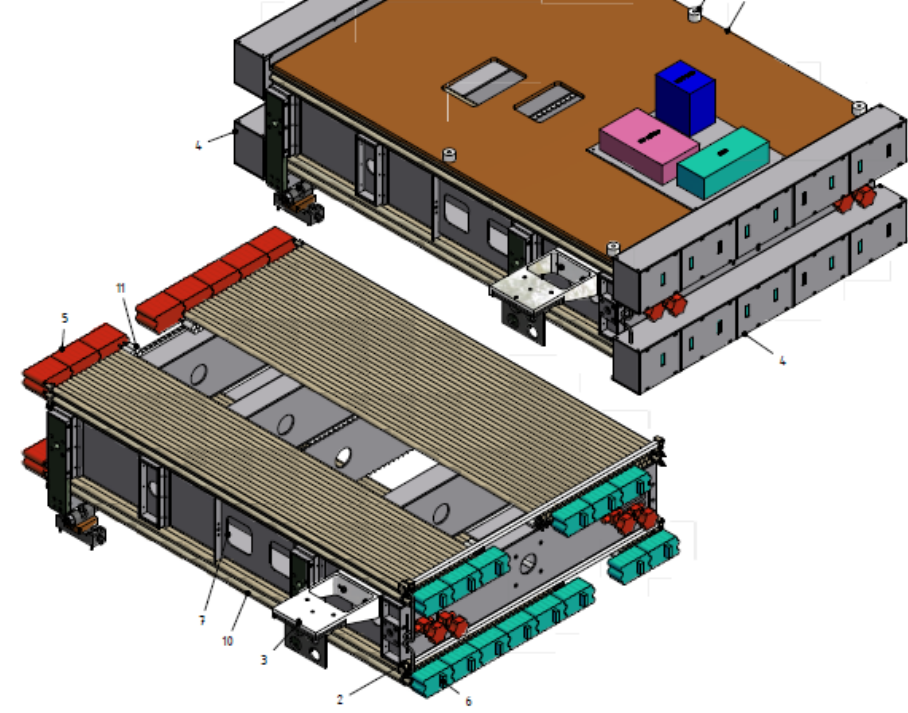}
	\end{subfigure}
	\caption{Illustration of the location of the new sMDT chambers (design drawings right) in the barrel middle layer of the ATLAS muon spectrometer (top; in red).}
	\label{fig::location}
\end{figure}

The Monitored Drift Tube (MDT) chambers of the ATLAS muon spectrometer~\cite{ATLAS} demonstrated that they provide very precise and robust tracking over large areas. Goals of ATLAS muon detector upgrades are to increase the acceptance for precision muon momentum measurement and triggering and to improve the rate capability of the muon chambers in the high-background regions when the LHC luminosity increases.
Small-diameter Muon Drift Tube (sMDT) chambers have been developed for these purposes~\cite{sMDT}. With half the drift-tube diameter of the MDT chambers and otherwise unchanged operating parameters, sMDT chambers share the advantages with the MDTs, but have an order of magnitude higher rate capability~\cite{sMDT} and can be installed in detector regions where MDT chambers do not fit in.
The construction of twelve chambers for the feet regions (see \fig{fig::location}) of the ATLAS detector is currently ongoing with the goal to install them in the winter shutdown 2016/17 of the LHC.
The purpose is to increase the acceptance for three-point measurement in these regions which will substantially improve the momentum resolution.

We report here about the construction of these chambers which started in the fall of 2014.
The chamber design and construction procedures have been optimized for mass production while providing high mechanical accuracy of better than 10~$\mu$m in the sense wire positioning.
The sMDT drift tubes are standard industrial aluminium tubes with a diameter of 15~mm and a wall thickness of 0.4~mm.

\begin{figure}[H]
	\centering
	\includegraphics[width=0.5\textwidth]{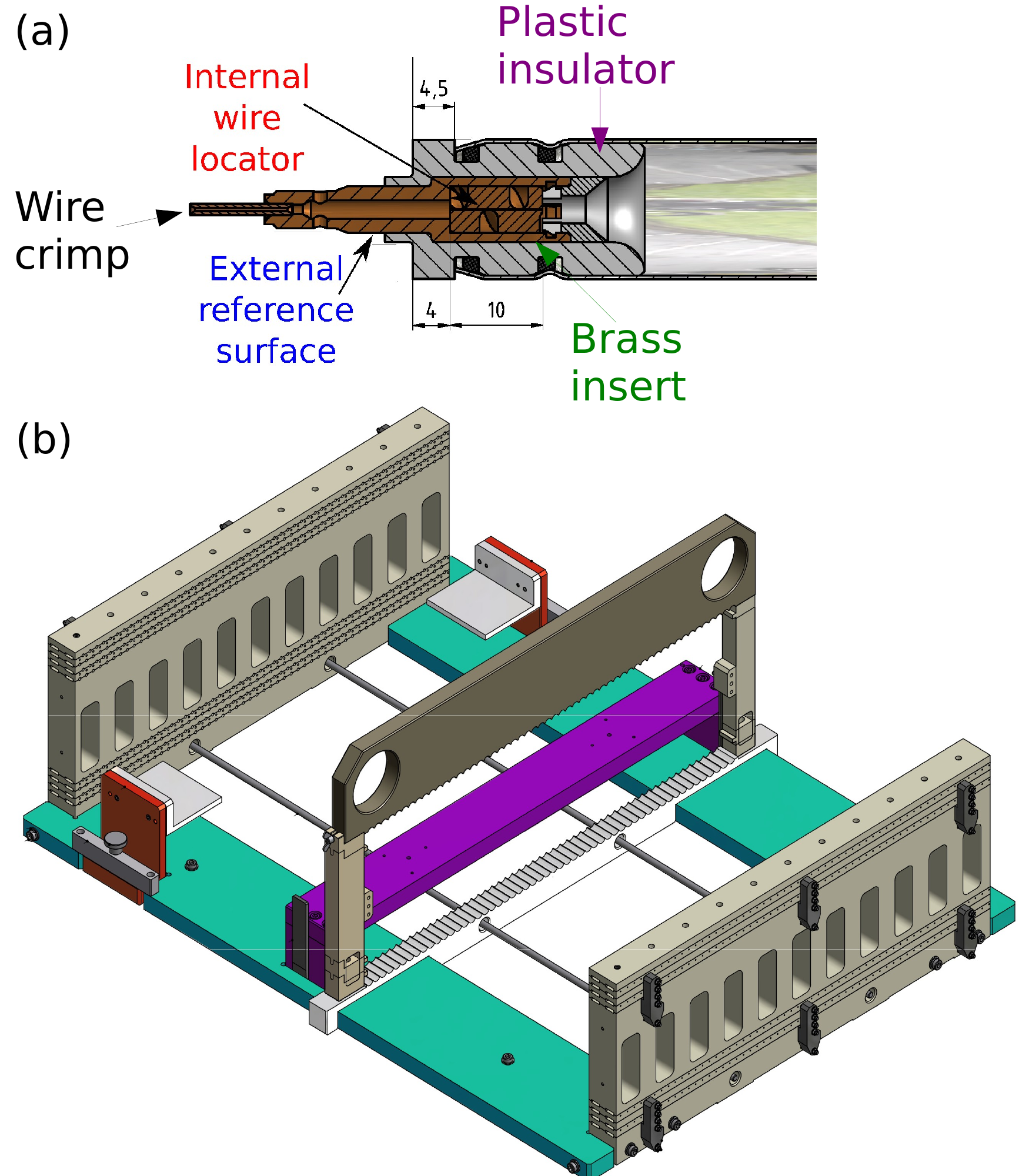}
	\caption{(a) The endplugs of the drift tubes consist of a plastic insulator (grey) and a brass insert (bown) which positions the sense wire with better than 5~$\mu$m accuracy with respect to a cylindrical reference surface on the outside of the tube~\cite{Pisa}. 
	(b) The reference surfaces are used to position the tubes during chamber assembly in holes in a precisely machined jig with the complete wire grid of the chamber at each tube end~\cite{Pisa}.}
	\label{fig::end_cap_chamber}
\end{figure}

\begin{figure}[H]
	\centering
	\begin{subfigure}[b]{0.5\textwidth}
		\centering
		\includegraphics[width=1\textwidth]{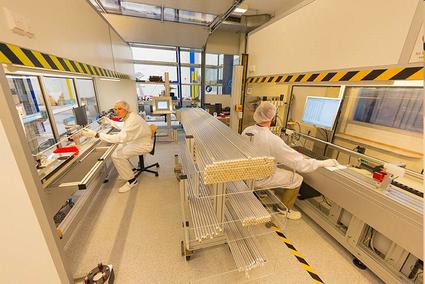}
		\caption{}
		\label{fig::clean1}
	\end{subfigure}
	\begin{subfigure}[b]{0.5\textwidth}
		\centering
		\includegraphics[width=1.0\textwidth]{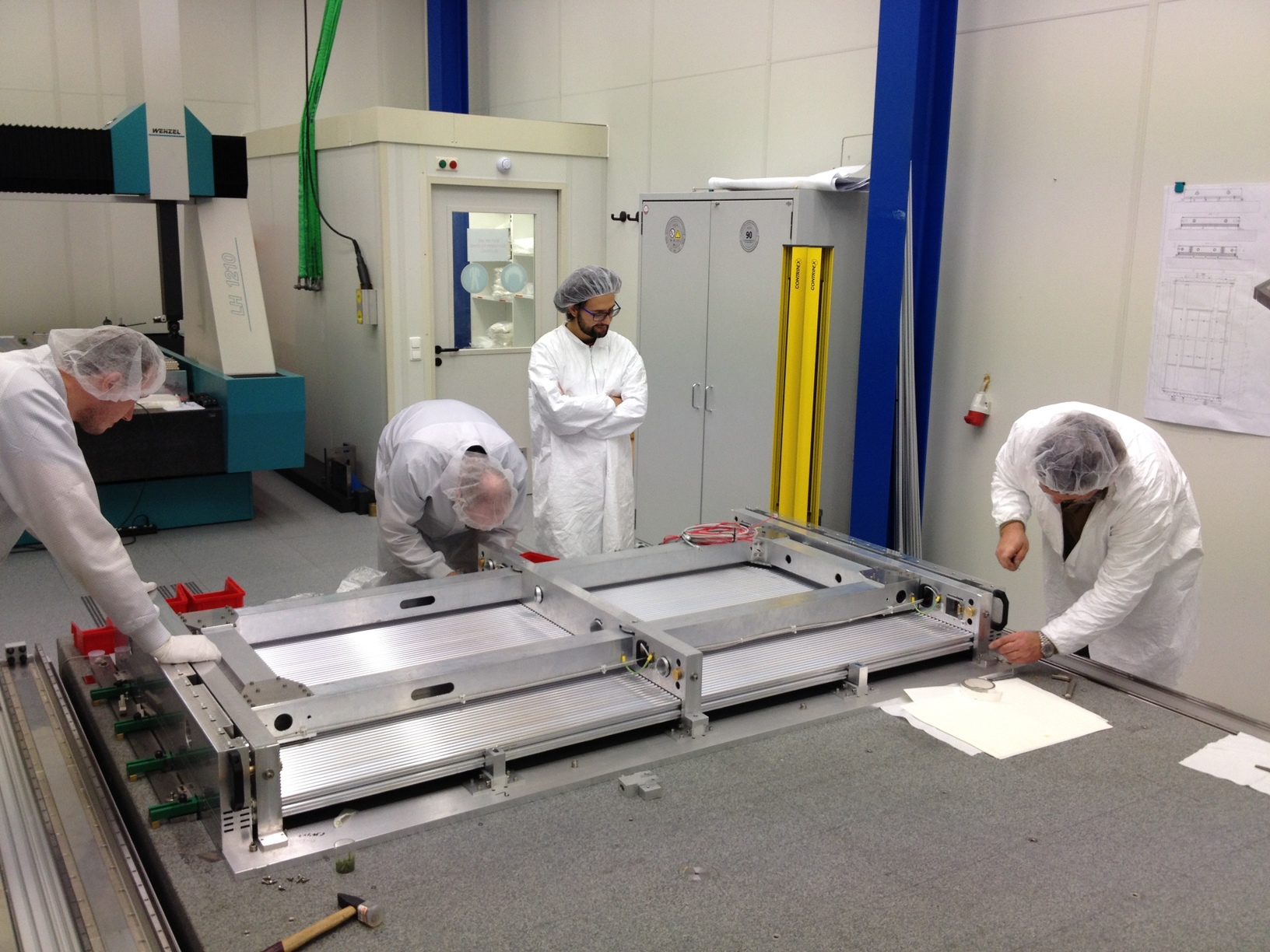}
		\caption{}
		\label{fig::clean2}
	\end{subfigure}
	\begin{subfigure}[b]{0.5\textwidth}
		\centering
		\includegraphics[width=1.0\textwidth]{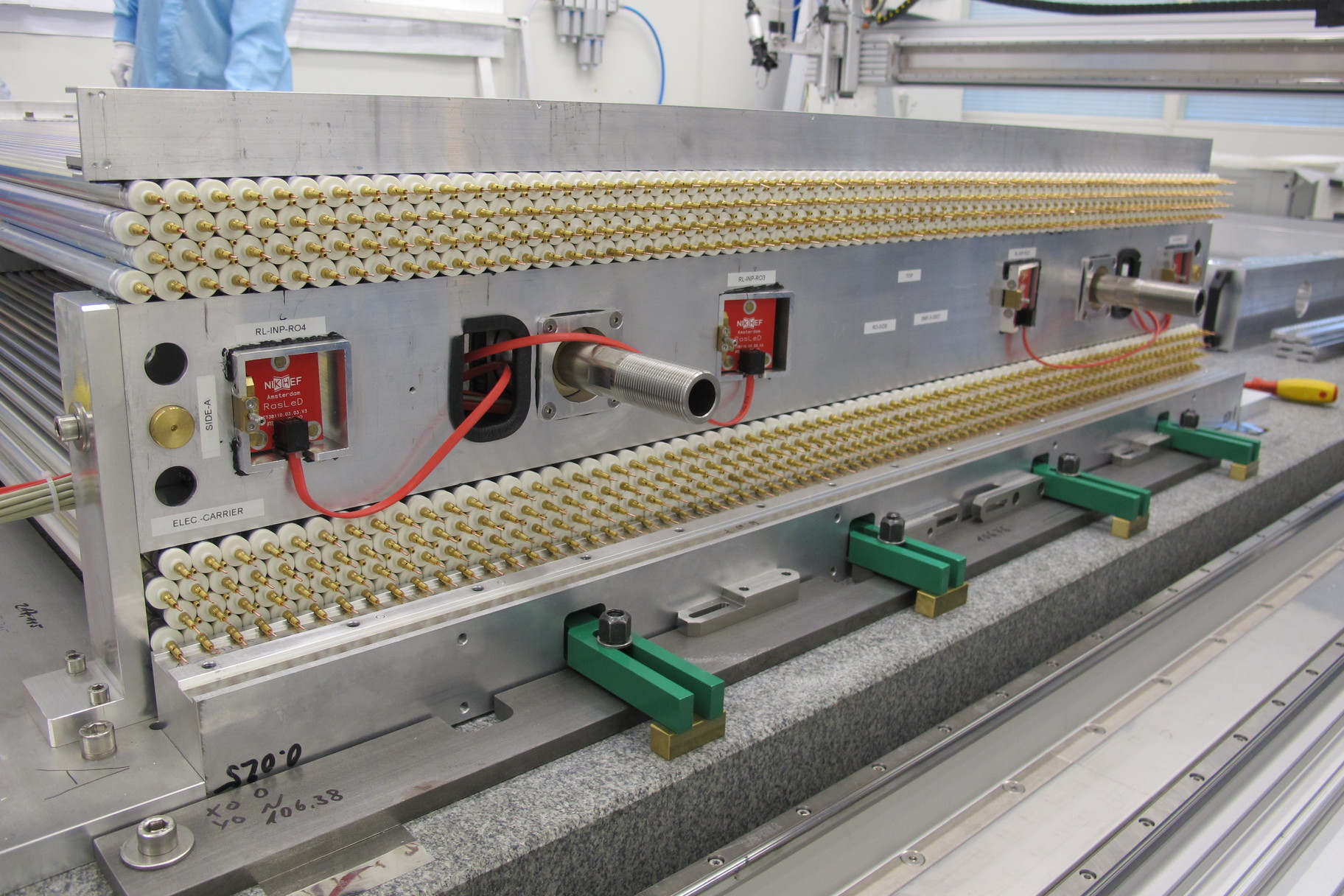}
		\caption{}
		\label{fig::clean3}
	\end{subfigure}
	\caption{(\subref{fig::clean1}) Semi-automatized drift tube assembly in a climatized clean room. On the same room the drift-tube tests are performed (see text).
	(\subref{fig::clean2}) Assembly of a sMDT chamber with the jigging described in the text and in Fig. 2 on a flat granite table in a climatized clean room. (\subref{fig::clean3})  Assembled chamber, still on the assembly table with two multilayers with four drift-tube layers each~\cite{Pisa}.}
	\label{fig::clean_room}
\end{figure}

\begin{figure}[H]
	\centering
	\begin{subfigure}[b]{0.25\textwidth}
		\centering
		\includegraphics[width=1\textwidth]{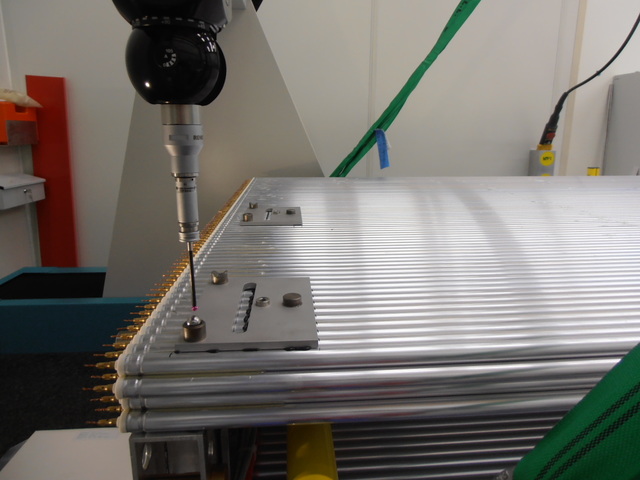}
		\caption{}
		\label{fig::meas1}
	\end{subfigure}
	\begin{subfigure}[b]{0.5\textwidth}
		\centering
		\includegraphics[width=1.0\textwidth]{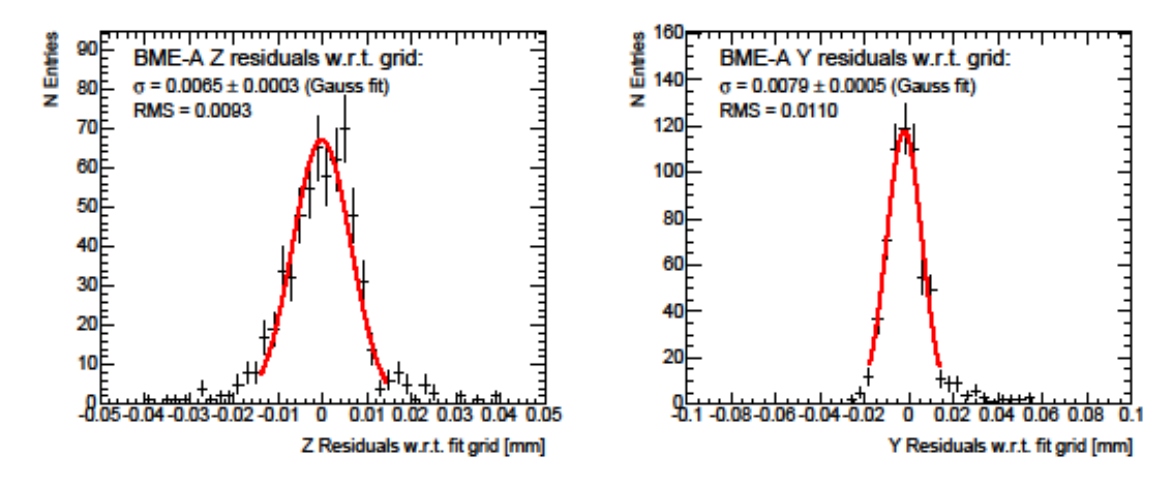}
		\caption{}
		\label{fig::meas2}
	\end{subfigure}
	\caption{(\subref{fig::meas1}) Measurement of wire and alignment sensor positions in a sMDT chamber on a coordinate measuring machine.
	(\subref{fig::meas2}) A sense wire positioning accuracy of better than 10~$\mu$m in both transverse coordinates z and y has been achieved (see the residual distribution with respect to the ideal wire grid on the left).
}
	\label{fig::clean_room}
\end{figure}

First the drift tubes are assembled with a semi-automatized wiring machine (see \fig{fig::clean1}).
The wire tension, the gas leak rate and the leakage current under high voltage are measured for every tube before it is assembled in a chamber.
The failure rate is only about 1\%.
About 100 tubes can be produced and tested per day with this setup.
The chambers are assembled by inserting the tubes with the reference surfaces on the endplugs into the jigs at each chamber end which define the wire grid as described in \fig{fig::end_cap_chamber} and glueing them together and to a spacer and support frame (see \fig{fig::location} and \fig{fig::clean2}).
A complete chamber can be assembled within one working day.
Before mounting the modular gas distribution system and the readout and high-voltage distribution boards~\cite{sMDT}, the wire grid and the geometry of the assembled chamber is reconstructed with a few micron precision by measuring the reference surfaces of the endplugs at both chamber ends on a coordinate measuring machine (see \fig{fig::meas1}).
An overall sense wire positioning accuracy in a chamber of better than 10~$\mu$m (see \fig{fig::meas2}).
The positions of the optical alignment sensors on the chambers are also measured in this way with respect to the wire grid.

Finally we report about the test results of the completed chambers in a cosmic ray teststand.

\end{document}